\documentclass[a4,prb,onecolumn,showpacs]{revtex4}
\usepackage{epsfig}
\usepackage{amsmath}

\def\e{\begin{equation}}
\def\f{\end{equation}}
\def\=#1{\overline{\overline{#1}}}
\def\_#1{{\bf #1}}
\def\^#1{\hat{#1}}
\def\o{\omega}
\def\E{\varepsilon}
\def\M{\mu}
\def\.{\cdot}
\def\x{\times}
\def\xx{{\ss{\x\atop\x}}}

\def\l#1{\label{eq:#1}}
\def\r#1{(\ref{eq:#1})}
\def\d{\nabla}

\def\ds{\displaystyle}
\def\ss{\scriptstyle}

\def\l#1{\label{eq:#1}}
\def\r#1{(\ref{eq:#1})}

\renewcommand{\Re}{\mathop{\rm Re}\nolimits}
\renewcommand{\Im}{\mathop{\rm Im}\nolimits}

\begin{document}

\title{Perfect lensing with phase conjugating surfaces:\\ Towards practical realization}

\author{Stanislav Maslovski$^1$}

\author{Sergei Tretyakov$^2$}

\affiliation{$^1$Departmento de Engenharia Electrot\'ecnica, Instituto de Telecomunica\c c\~oes - Universidade de Coimbra, P\'olo II, 3030 Coimbra, Portugal\\ $^2$Aalto University, School of Electrical Engineering,
P.O. 13000, FI-00076 Aalto, Finland\\
{\rm E-mails: stas@co.it.pt,
sergei.tretyakov@aalto.fi}}

\date{\today}

\begin{abstract}
  It is theoretically known that a pair of phase conjugating surfaces
  can function as a perfect lens, focusing propagating waves and
  enhancing evanescent waves. However, the known experimental
  approaches based on thin sheets of nonlinear materials cannot fully
  realize the required phase conjugation boundary condition. In this
  paper we show that the ideal phase conjugating surface is in
  principle physically realizable and investigate the necessary
  properties of nonlinear and nonreciprocal particles which can be
  used to build a perfect lens system. The physical principle of the
  lens operation is discussed in detail and directions of possible
  experimental realizations are outlined.
\end{abstract}


\maketitle

\section{Introduction}

The {\itshape perfect lens}~\cite{Pendry0} is a device which focuses
the field of a point source into a point, that is, the perfect lens
focuses both propagating and evanescent fields. It is
known~\cite{Pendry0} that a planar slab of the ideal Veselago
medium~\cite{Veselago} with the relative permittivity and permeability both
equal to $-1$ has the perfect-lens properties because of the negative
refraction phenomenon and the excitation of coupled surface
plasmon-polaritons at the slab interfaces. Practical realization of
such double-negative (DNG) materials is, however, a significant
challenge, especially at optical frequencies, and, clearly, any
realization will suffer from some imperfections. For example,
metal-dielectric metamaterials have relatively high ohmic losses that
are responsible for nonvanishing imaginary parts in the constitutive
parameters of such volumetric artificial media.

There is, however, a possibility for a different realization of a
perfect lens that does not require any volumetric metamaterials.
Indeed, the physical effects necessary for perfect lensing happen at
the lens interfaces, and not within the metamaterial
volume. Therefore, if one realizes a {\em metasurface} at which the
incident waves refract negatively, then a parallel pair of such planar
sheets will mimic the operation of the Veselago lens for the {\em
  propagating} plane waves. Moreover, if this metasurface supports
surface modes (surface plasmon-polaritons) within a wide range of the
tangential propagation factors $k_{\rm t} > k_0$ (where $k_0$ is the
free space wavenumber), then also the impinging {\em evanescent} plane
waves will interact resonantly with the sheets and will be tunneled
through the lens with an enhanced amplitude, due to the
electromagnetic coupling between the surface states excited on the
sheets. Such subwavelength imaging with {\em linear} plasmon-polariton
resonant grids was theoretically predicted and confirmed
experimentally in a number of
works.\cite{Maslovski_grid,Marques_SRR,Alitalo_cylinder,Alitalo_nanospheres,Maslovski_freqscan}

In 2003, we showed~\cite{JAP} that two parallel sheets with phase
conjugating boundary conditions for tangential fields on the two sides
of the sheets \e \_E_{\rm t+}=\_E_{\rm t_-}^*, \qquad \_H_{\rm
  t+}=\_H_{\rm t-}^* \l{BC0} \f have the necessary properties of the
perfect lens outlined above. In these conditions that are written for
the complex amplitudes of the time-harmonic fields (symbol $^*$
denotes complex conjugation operation) the indices $\pm$ indicate the
field values on the two sides of an infinitely thin phase conjugating
sheet.

Obviously, boundary conditions \r{BC0} cannot be realized using linear
materials, and in the same paper~\cite{JAP} the use of three-wave
mixing in a nonlinear layer was proposed as an approach to realization
of this effect. Phase conjugation and ``time-reversal'' devices were
studied also earlier for other applications. It is interesting that in
the same year (2003) an experimental microwave realization of a phase
conjugating layer was published~\cite{Itoh} independently from our
work.\cite{JAP} Later, the concept of perfect lensing based on two
nonlinear sheets was studied theoretically in Ref.~\onlinecite{Pendry}.
Alternative experimental realizations of nonlinear negative refraction
effect were published in Refs.~\onlinecite{Fusco,Shvets}.

However, in known devices based on antenna arrays with
mixers,\cite{Itoh} or sheets of nonlinear dielectrics,\cite{Pendry}
or arrays of only electric or only magnetic particles with nonlinear
insertions,\cite{Shvets} the phase conjugated (``time-reversed'')
products create waves propagating symmetrically to the both sides of
the sheet, i.e., there appears a retrodirected wave propagating back
to the source. While the perfect lens operation can be theoretically
approached even in such systems if the amplitudes of the nonlinear
products tend to infinity in the assumption of nonphysical infinitely
strong external pumping,\cite{Pendry} the ideal phase conjugating
boundary conditions~\r{BC} cannot be realized within this scenario.

In this paper we discuss the physical meaning of the ideal
complex-conjugation boundary conditions \r{BC0} and outline possible
approaches for realization of such surfaces, which would potentially
lead to creation of super-resolution lenses. The paper is organized as
follows. In Section~\ref{physics} we consider physical processes
taking place at a phase conjugating boundary and demonstrate that
such boundary may be equivalently represented with pairs of electric
and magnetic surface currents reacting (nonlinearly) to the applied
magnetic and electric fields. In Section~\ref{bian} a realization of
the phase conjugating boundary with an array of bi-anisotropic
inclusions is proposed and studied and the necessary conditions on nonlinear
susceptibilities of the inclusions are established. In
Section~\ref{secdesign} a possible microwave design of such inclusions
is proposed.

\section{\label{physics}The physical meaning of the
  complex-conjugating boundary conditions}

\subsection{Complex-conjugating boundary and perfect lens}

\begin{figure}
\centering
\epsfig{file=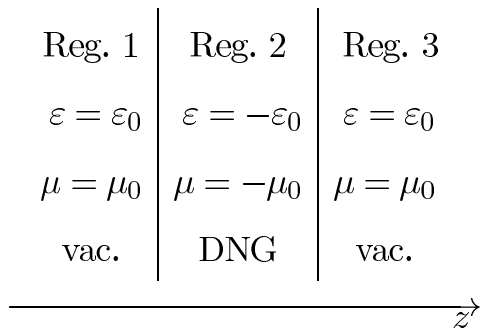,width=0.35\textwidth}
\caption{An ideal Veselago lens: A planar slab of a double-negative (DNG) material
with the medium parameters $\E=-\E_0$ and $\M=-\M_0$ in free space.}
\label{fig1}
\end{figure}

Let us start from outlining the idea from our paper in Ref.~\onlinecite{JAP}.
Consider an ideal Veselago lens depicted in Figure~\ref{fig1}.  Let
the relative permittivity and permeability of the medium surrounding
the lens be equal to~$1$ and the relative parameters of the lens
material to~$-1$ at the working frequency~$\o$, respectively.
At the lens interfaces the tangential components of the fields satisfy
the usual Maxwellian continuity conditions.  One may notice that in
this system the only difference between the time-harmonic [the time
dependence is of the form $\exp(+j\o t)$] field equations in the
Veselago slab (region 2) \e \d\x\_E =j\o\M_0\_H, \qquad \d\x\_H =
-j\o\E_0\_E \f and the analogous equations in the free-space regions
is the sign in front of the imaginary unit. A substitution \e \_E_{\rm
  (old)}, \_H_{\rm (old)} \Rightarrow C\_E_{\rm (new)}^*, C\_H_{\rm
  (new)}^* \l{subst} \f ($C$ is an arbitrary constant; here and
thereafter $^*$ denotes the complex conjugation) into the field
equations in region 2 reformulates the problem in terms of the new
field vectors in which the field equations become the same in all
three regions: \e \d\x\_E = -j\o\M_0\_H,\qquad \d\x\_H = j\o\E_0\_E,
\f that are simply the Maxwell equations in free space. The boundary
conditions on the two interfaces, however, are no more the standard
continuity conditions, but they involve complex conjugation: \e
\_E_{\rm t_{(1,3)}}=C\_E_{\rm t_{(2)}}^*, \qquad \_H_{\rm
  t_{(1,3)}}=C\_H_{\rm t_{(2)}}^*. \l{BCC} \f

The constant $C$ describes the ``transformation efficiency'' of the
nonlinear surface which transforms fields into the complex-conjugate
state. In the known experimental realizations of phase conjugation in
electrically thin layers (e.g., Ref.~\onlinecite{Shvets}), the
efficiency has been rather small. However, by choosing a small value
of $C$ in \r{BCC}, we arrive at a structure with asymmetric properties
with respect to the two sides of the surface. Indeed, complex
conjugating and dividing \r{BCC} by $C^*$, we see that in this case
the weak fields inside the lens should be enhanced by the surface in
the same rate as they are suppressed when the surface is excited from
outside. For this reason we will concentrate on the simplest choice of
$C=1$, as in Ref.~\onlinecite{JAP}, described by the boundary conditions \e
\_E_{\rm t_{(1,3)}}=\_E_{\rm t_{(2)}}^*, \qquad \_H_{\rm
  t_{(1,3)}}=\_H_{\rm t_{(2)}}^*. \l{BC} \f In this case the
complex-conjugating surface has symmetric properties with respect to
its two sides, and for the ideal lens operation the amplitudes of the
field should not change across the sheets.

Now it becomes evident that the problem involving an ideal Veselago
slab is mathematically equivalent to the problem dealing with a pair
of conjugating surfaces in free space, provided that the field sources
are outside of region 2. Therefore, in the latter system the field
solutions are the same as in the Veselago slab, and because of this
the physical phenomena taking place at the interfaces of region 2 are
also the same: The propagating plane waves are refracted negatively at
the interfaces, and the evanescent modes are enhanced due to the
excitation of coupled surface plasmon-polariton pairs. In this regard,
a pair of phase conjugating planes is indistinguishable from a perfect
lens proposed by Pendry.\cite{Pendry0}
In what follows we concentrate on physical properties of such
phase conjugating sheets.

\subsection{Plane-wave propagation across the phase conjugating
sheet}

Let us consider a single phase conjugating surface located at $z =
0$. We decompose the tangential electric and magnetic fields into
plane waves at both sides of the surface:
\begin{eqnarray}
\_E_{\rm t}(x,y)\Big|_{z = \pm 0} &=& {1\over (2\pi)^2} \int\int
\_E_{\rm t}(k_x,k_y)\Big|_{z = \pm 0}
e^{-j(k_x x + k_y y)}\,dk_x dk_y,\\
\_H_{\rm t}(x,y)\Big|_{z = \pm 0} &=& {1\over (2\pi)^2} \int\int
\_H_{\rm t}(k_x,k_y)\Big|_{z = \pm 0} e^{-j(k_x x + k_y y)}\,dk_x
dk_y.
\end{eqnarray}

It is easy to see that the boundary conditions \r{BC} require that
the plane-wave components satisfy
\begin{eqnarray}
\_E_{\rm t}(k_x,k_y)\Big|_{z = +0} &=& \_E_{\rm t}^{*}(-k_x,-k_y)\Big|_{z = - 0},\\
\_H_{\rm t}(k_x,k_y)\Big|_{z = +0} &=& \_H_{\rm
t}^{*}(-k_x,-k_y)\Big|_{z = - 0}.
\end{eqnarray}
From these relations we immediately realize that the propagating
modes refract negatively at the conjugating interface due to the
change in the sign of {\em the tangential} component of the wave
vector $\_k_{\rm t} = k_x\_x_0 + k_y\_y_0$. This is very different
from the case of the same refraction at an interface with a
double-negative medium where {\em the
  normal} component of the wave vector changes sign.

What about the evanescent modes? It can be shown that they are at
resonance with the phase conjugating surface so that the strong
(theoretically infinite) reflection takes place at the surface, although the
field transformation in the sheets (the complex conjugate operation) does not amplify the fields. The
details are given in Ref.~\onlinecite{JAP}; here we will just try to convince
the reader with the following simple physical argument.

Let us consider a single plane-wave component interacting with a
single phase conjugating surface in free space. The plane wave is
incident from the half space $z < 0$. On both sides of the phase
conjugating sheet the tangential fields satisfy the usual relation
between the fields in a free-space plane wave (the following equations
are valid for TM or TE waves separately): \e \_E_{\rm
  t}(k_x,k_y)\Big|_{z=\pm 0} = -Z_{\rm TM,TE}\ \_z_0\x\_H_{\rm
  t}(k_x,k_y)\Big|_{z=\pm 0}. \l{impe} \f Here the free-space wave
impedances for TM- and TE-polarized waves read \e \l{Ztm} Z_{\rm
  TM}=\eta_0\sqrt{1 - {k_{\rm t}^2\over k_0^2}},\qquad Z_{\rm
  TE}=\eta_0{1\over \sqrt{1 - {k_{\rm t}^2\over k_0^2}}},\f where
$k_0=\omega/c$ is the free-space wavenumber, $\eta_0 =
\sqrt{\M_0/\E_0}$, and $k_{\rm t}^2=k_x^2+k_y^2$.  But on the other
hand the tangential fields satisfy also the boundary condition
\r{BC}. Together with the impedance relation \r{impe} this leads to
\begin{align} \_E_{\rm
t}(k_x,k_y)\Big|_{z=-0} &=  \_E^*_{\rm
t}(-k_x,-k_y)\Big|_{z=+0} \nonumber \\
&= -Z^*_{\rm TM,TE}\ \_z_0\x\_H^*_{\rm
t}(-k_x,-k_y)\Big|_{z=+0} \nonumber \\
&= -Z^*_{\rm TM,TE}\ \_z_0\x\_H_{\rm
t}(k_x,k_y)\Big|_{z=-0}.
\end{align}
Thus, a wave incident on the sheet from one side ``sees'' the surface
impedance which equals to the complex conjugate of the wave impedance
in free space. The reflection coefficient reads
\e \l{refl}
R_{\rm TM,TE}={Z^*_{\rm TM,TE}-Z_{\rm TM,TE}\over
{Z^*_{\rm TM,TE}+Z_{\rm TM,TE}}}.
\f

If the wave is a propagating wave, that is, $|\_k_{\rm t}| \equiv k_{\rm t} \equiv
\sqrt{k_x^2+k_y^2}< k_0 \equiv \omega/c$, then its wave impedance is
a real number, and the reflection coefficients equal zero.
This proves that the propagating modes experience negative
refraction without any reflection.

If the wave is evanescent, that is, $k_{\rm t} \equiv
\sqrt{k_x^2+k_y^2}> k_0$, the wave impedances \r{impe} are purely
imaginary and the reflection coefficient is infinite. The transmission
coefficient is also infinite due to the boundary condition
\r{BCC}. This is obviously a resonant condition which can be also
understood as a condition for existence of a surface mode.\cite{JAP}
Note that this resonant condition holds for all values of
the tangential wave number $k_{\rm t}>k_0$, which is the condition for
perfect lensing of all evanescent field components.

The physical reason of such a resonance is rather simple. Consider,
for instance, the waves of TM polarization.  The characteristic
impedance of an evanescent TM wave \r{Ztm} can be written as $Z_{\rm
  TM} = -{j\eta_0\alpha/k_0}$, where $\alpha$ is the decay factor:
$\alpha = \sqrt{k_{\rm t}^2 - k_0^2}$. We conclude that a TM
evanescent wave has capacitive characteristic impedance. Due to the
conjugating interface, the same impedance of the matching TM wave
behind the interface (at $z = +0$) is seen as inductive in front of
the interface (at $z = -0$), which reminds of a ``connection'' of
two reactances of the opposite character in a parallel oscillatory
circuit.  Probably, A. Al\`u and N. Engheta~\cite{Engheta} were the
first to identify and explain the resonance of the same nature that
happens at the border of a double-positive and a double-negative
material.

Therefore, when an incident evanescent wave excites a conjugating
surface, it resonantly excites a surface mode that matches its
transversal propagation factor and this results in very strong
(theoretically infinite) reflected and transmitted waves at the
surface. The strong reflection was also found to be the key to
sub-wavelength imaging in a pair of parametrically pumped nonlinear
{\em nonmagnetic} sheets studied in Ref.~\onlinecite{Pendry} The above
discussion shows that in a metasurface realizing the boundary
conditions \r{BC} the reflection and transmission coefficients for
evanescent modes tend to infinity due to a high-quality resonance
(theoretically, with an infinite quality factor), while, in
Ref.~\onlinecite{Pendry}, the surface itself parametrically amplifies the fields.

\subsection{\label{currents}Equivalent surface currents on the phase conjugating sheet}

One may notice that the boundary conditions \r{BC} imply discontinuity
of both tangential electric and magnetic fields across the phase
conjugating plane. The jumps of the fields can be expressed as
follows:
\begin{eqnarray}
\_E_{\rm t}(x,y)\Big|_{z=+0} - \_E_{\rm t}(x,y)\Big|_{z=-0} &=& -2j\Im[\_E_{\rm t}(x,y)]\Big|_{z=-0},\\
\_H_{\rm t}(x,y)\Big|_{z=+0} - \_H_{\rm t}(x,y)\Big|_{z=-0} &=&
-2j\Im[\_H_{\rm t}(x,y)]\Big|_{z=-0}\l{EH_jumps}.
\end{eqnarray}
These jumps are related to the equivalent
magnetic and electric surface currents that exist on the surface:
\begin{eqnarray}
\l{surfcur1}
\_z_0\x\_J_{\rm m}(x,y) &=& \_E_{\rm t}(x,y)\Big|_{z=+0} - \_E_{\rm
  t}(x,y)\Big|_{z=-0}\nonumber  \\ & = & -2j\Im[\_E_{\rm t}(x,y)]\Big|_{z=-0}=2j\Im[\_E_{\rm t}(x,y)]\Big|_{z=+0},\\
\l{surfcur2} -\_z_0\x\_J_{\rm e}(x,y) &=& \_H_{\rm
t}(x,y)\Big|_{z=+0} - \_H_{\rm t}(x,y)\Big|_{z=-0}\nonumber \\ & = &
-2j\Im[\_H_{\rm t}(x,y)]\Big|_{z=-0}=2j\Im[\_H_{\rm
t}(x,y)]\Big|_{z=+0}.
\end{eqnarray}

Let us stress already at this point that the
relations~\r{surfcur1}--\r{surfcur2} are the only physical conditions
one has to satisfy in a subwavelength imaging device based on phase
conjugating sheets. Nothing more is required!  Essentially, these
relations tell us that it is enough in practice to realize a
metasurface that reacts with certain magnetic and electric surface
currents to the imaginary parts of the tangential electric and
magnetic fields {\em at a given side} of the surface. This also
shows that there is no need for extra-strong pumping or
super-efficient nonlinear conversion as in Ref.~\onlinecite{Pendry}. Moreover,
from \r{surfcur1}--\r{surfcur2} we observe that the induced
{\itshape electric} current should be proportional to the {\itshape
  magnetic} field on the sheet (more precisely, to its imaginary
part). Likewise, the induced {\itshape magnetic} current is
proportional to the {\itshape electric} field. This is quite different
from the known approaches based on layers of nonlinear
dielectrics or magnetics.\cite{Pendry,Shvets}

Let us decompose each of these surface currents into a sum of two
currents: $\_J_{\rm e} = \_J_{\rm e}^{(1)} + \_J_{\rm e}^{(2)}$,
$\_J_{\rm m} = \_J_{\rm m}^{(1)} + \_J_{\rm m}^{(2)}$, where
\begin{eqnarray}
\l{huygens1}
  -\_z_0\x\_J_{\rm e}^{(1)} = -\_H_{\rm t}\Big|_{z = -0}&,& \quad
  \_z_0\x\_J_{\rm m}^{(1)} = -\_E_{\rm t}\Big|_{z = -0},\\
\l{huygens2}
  -\_z_0\x\_J_{\rm e}^{(2)} = \_H_{\rm t}\Big|_{z = +0}&,& \quad
  \_z_0\x\_J_{\rm m}^{(2)} = \_E_{\rm t}\Big|_{z = +0}.
\end{eqnarray}

One can see that the pair of the surface currents $\_J_{\rm
e}^{(1)}$, $\_J_{\rm m}^{(1)}$ is essentially an equivalent Huygens
source defined at the plane $z = -0$.  In the half-space $z > 0$
this source produces the field which is the negate of the field that
the external sources located at $z < 0$ induce in the half-space $z
> 0$ (the negation is due to the minus signs in the right-hand side
of \r{huygens1}). Thus, the physical role of the currents $\_J_{\rm
  e}^{(1)}$, $\_J_{\rm m}^{(1)}$ when concerned with the half-space $z
> 0$ is to cancel the field incident from the other half-space. The
same holds for the other pair of currents $\_J_{\rm e}^{(2)}$,
$\_J_{\rm m}^{(2)}$ when concerned with the half-space $z < 0$.
These currents form a Huygens source defined at $z = +0$ plane. In
the half-space $z < 0$ they cancel (pay attention to the direction
of the normal!) the field produced by the external sources from the
$z > 0$ half-space.

From the other hand, the pair of currents $\_J_{\rm e}^{(1)}$,
$\_J_{\rm m}^{(1)}$ plays another role when concerned with the
half-space $z < 0$. Indeed, using the boundary conditions \r{BC} we
write \e -\_z_0\x\_J_{\rm e}^{(1)} = -\_H_{\rm t}^{*}\Big|_{z = +0},
\quad \_z_0\x\_J_{\rm m}^{(1)} = -\_E_{\rm t}^{*}\Big|_{z = +0}, \f
from which it is evident that these currents can be identified also
as an equivalent source located at the plane $z = +0$ that produces
at $z < 0$ the conjugated field of the sources located in the
half-space $z
> 0$. Respectively, $\_J_{\rm e}^{(2)}$, $\_J_{\rm m}^{(2)}$ produce
the conjugated field in the region $z > 0$.

To summarize, we have identified the following roles of the
currents:
\begin{itemize}
\item The pair $\_J_{\rm e}^{(1)}$, $\_J_{\rm m}^{(1)}$ cancels the
  field incident from $z < 0$ to the half-space $z > 0$ and creates
  the conjugated field of the sources from the half-space $z > 0$ in
  the half-space $z < 0$;
\item The pair $\_J_{\rm e}^{(2)}$, $\_J_{\rm m}^{(2)}$ cancels the
  field incident from $z > 0$ to the half-space $z < 0$ and creates
  the conjugated field of the sources from the half-space $z < 0$ in
  the half-space $z > 0$.
\end{itemize}

\section{\label{bian}Phase conjugating surface as an array of bi-anisotropic
  nonlinear inclusions}

From the above results we see that the ideal phase conjugating
surface should respond to the fields with both electric and magnetic
polarization. Dependence of the induced electric current on the
magnetic field and {\it vice versa} suggests that the structure
should have some magneto-electric coupling. In this section we
investigate if it is possible to realize the ideal phase conjugating
boundary conditions \r{BC} with a planar array of nonlinear
bi-anisotropic particles.

\subsection{\label{genreq}General requirements on susceptibilities of inclusions}

Let us first find out how the total induced electric and magnetic
surface current densities depend on the {\em incident} electric and
magnetic fields in the array plane. To do that, we consider an
isolated phase conjugating surface in the field of a {\em single} TM
(or TE) polarized plane electromagnetic wave (propagating or
evanescent). Taking into account the conjugating boundary condition
\r{BC} we may formally write the total tangential electric and
magnetic fields on both sides of the surface as
\begin{align}
\l{Einc1}
\_E_{\rm t}(x,y)\Big|_{z=-0} &= (1 + R_{\rm TM,TE})\_E^{\rm inc}_{\rm t}(x,y)\Big|_{z=0},\\
\_E_{\rm t}(x,y)\Big|_{z=+0} &= (1 + R_{\rm TM,TE}^*)\left(\_E^{\rm inc}_{\rm t}(x,y)\right)^*\Big|_{z=0},\\
\_H_{\rm t}(x,y)\Big|_{z=-0} &= (1 - R_{\rm TM,TE})\_H^{\rm inc}_{\rm
  t}(x,y)\Big|_{z=0},\\
\l{Hinc2}
\_H_{\rm t}(x,y)\Big|_{z=+0} &= (1 - R_{\rm TM,TE}^*)\left(\_H^{\rm inc}_{\rm t}(x,y)\right)^*\Big|_{z=0},
\end{align}
where the reflection coefficients $R_{\rm TM,TE}$ are given by
\r{refl}, from which we notice that $R_{\rm TM,TE}^* = -R_{\rm
  TM,TE}$. The above expressions hold for both propagating and
evanescent waves incident from the half space $z < 0$.

Therefore, from \r{surfcur1}--\r{surfcur2} and \r{Einc1}--\r{Hinc2} we obtain
\begin{eqnarray}
\l{JmEinc}
-\_z_0\x\_J_{\rm m}(x,y) &=& 2j\Im(\_E^{\rm inc}_{\rm t}(x,y))\Big|_{z=0}+2R_{\rm TM,TE}\Re(\_E^{\rm inc}_{\rm t}(x,y))\Big|_{z=0},\\
\l{JeHinc}
\_z_0\x\_J_{\rm e}(x,y) &=& 2j\Im(\_H^{\rm inc}_{\rm t}(x,y))\Big|_{z=0}-2R_{\rm TM,TE}\Re(\_H^{\rm inc}_{\rm t}(x,y))\Big|_{z=0}.
\end{eqnarray}
The addends on the right-hand side of \r{JmEinc}--\r{JeHinc} that are
proportional to the imaginary part of the incident field are relevant
for the propagating waves (for these waves $R_{\rm TM,TE} = 0$), while
for the evanescent waves the addends proportional to the real part of
the field are of the most importance, because $|R_{\rm
  TM,TE}|\rightarrow\infty$ for these modes. It is instructive to
compare these observations with the discussion in
Sec.~\ref{currents}. From~\r{surfcur1}--\r{surfcur2} it follows that
in terms of the total tangential fields at a given side of the phase
conjugating sheet, the conditions for both propagating and
evanescent waves are the same: Eqs.~\r{surfcur1}--\r{surfcur2} do
not distinguish these waves. Physically, this is because the
locations where one must {\it measure} the fields and where one must
{\it
  create} the surface currents {\it are not at the same point}, if one
wants to approach a design directly suggested by these equations.
Indeed, the mathematical form of Eqs.~\r{surfcur1}--\r{surfcur2}
demands that the field values must be taken at a point slightly
displaced off the surface $z = 0$. On the contrary, in an array of
particles (considered here as point objects) the equivalent surface
currents depend only on the fields {\it in
  the array plane} and, as we will see soon, the required
reaction to this field happens to be different for the two types of
waves.

A related observation is that in a realistic structure, e.g., an array of
nonlinear polarizable bi-anisotropic inclusions, it is not the
incident field, but the {\em local} field $\_E_{\rm t}^{\rm loc}$,
$\_H_{\rm t}^{\rm loc}$, that excites each and every inclusion in the
structure. The latter has a contribution from the secondary field of
the induced currents. We may therefore write for the signals at the
frequency of the incident wave:
\begin{align}
\l{eloc}
\_E^{\rm loc}_{\rm t}(x,y) &= \_E_{\rm t}^{\rm inc}(x,y) + \=\beta_{\rm
  ee}\.\_J_{\rm e}(x,y),\\
\l{hloc}
\_H^{\rm loc}_{\rm t}(x,y) &= \_H_{\rm t}^{\rm inc}(x,y) + \=\beta_{\rm
  mm}\.\_J_{\rm m}(x,y),
\end{align}
where $\=\beta_{\rm ee,mm}$ are the so-called interaction dyadics. For
arbitrary distributed currents these dyadics are understood as
operators acting on the currents. However, for the following it is
enough to consider the currents of the form $\_J_{\rm e,m}(x,y) =
\_J^{\rm e,m}_{\_k_{\rm t}}\exp(-j\_k_{\rm t}\.\_r) + \_J^{\rm
  e,m}_{-\_k_{\rm t}}\exp(j\_k_{\rm t}\.\_r)$. In this case, in order
for \r{eloc}--\r{hloc} to hold in a simple dyadic sense, the
interaction dyadics must satisfy $\=\beta_{\rm
  ee,mm}\equiv\=\beta_{\rm ee, mm}(\_k_{\rm t})=\=\beta_{\rm
  ee,mm}(-\_k_{\rm t})$, i.e., the lattice (not the particles!) must
have a center of symmetry. There are no cross-terms in
\r{eloc}--\r{hloc} because the tangential magnetic (electric) field of
an array of tangential electric (magnetic) dipoles vanishes in the
plane of the array.

Additionally, the induced electric and magnetic currents must be
sensitive to the phase of the external field, because the
conjugating boundary reacts differently to the real and imaginary
parts of the tangential electric and magnetic fields. Therefore, the
inclusions must react differently to the corresponding components of
the local fields. Based on the above discussion we may write
\begin{align}
 \l{elepolar} \_J_{\rm e} &= \=\alpha_{\rm ee}^{\rm re}\.\Re(\_E^{\rm
    loc}_{\rm t}) + j\=\alpha_{\rm ee}^{\rm im}\.\Im(\_E^{\rm
    loc}_{\rm t}) + \=\alpha_{\rm em}^{\rm re}\.\Re(\_H^{\rm loc}_{\rm
    t}) + j\=\alpha_{\rm em}^{\rm im}\.\Im(\_H^{\rm loc}_{\rm t}),\\
 \l{magpolar} \_J_{\rm m} &= \=\alpha_{\rm me}^{\rm re}\.\Re(\_E^{\rm
    loc}_{\rm t}) + j\=\alpha_{\rm me}^{\rm im}\.\Im(\_E^{\rm
    loc}_{\rm t}) + \=\alpha_{\rm mm}^{\rm re}\.\Re(\_H^{\rm loc}_{\rm
    t}) + j\=\alpha_{\rm mm}^{\rm im}\.\Im(\_H^{\rm loc}_{\rm t}),
\end{align}
where $\=\alpha_{\rm ee,mm,me,em}^{\rm re,im}$ are the dyadic polarizabilities
to the real and imaginary components of the local fields.

Substituting \r{magpolar}--\r{elepolar} into \r{eloc}--\r{hloc} we obtain
\begin{multline}
  \l{einc2} \_E^{\rm inc}_{\rm t} = (\=I_{\rm t} - \=\beta_{\rm
    ee}\.\=\alpha_{\rm ee}^{\rm re})\.\Re(\_E^{\rm loc}_{\rm t}) +
  j(\=I_{\rm t}-\=\beta_{\rm ee}\.\=\alpha_{\rm ee}^{\rm im})\.\Im(\_E^{\rm
    loc}_{\rm t}) \\- \=\beta_{\rm ee}\.\=\alpha_{\rm em}^{\rm
    re}\.\Re(\_H^{\rm loc}_{\rm
    t}) - j\=\beta_{\rm ee}\.\=\alpha_{\rm em}^{\rm im}\.\Im(\_H^{\rm
    loc}_{\rm t}),
\end{multline}
\begin{multline}
  \l{hinc2} \_H^{\rm inc}_{\rm t} = (\=I_{\rm t} - \=\beta_{\rm
    mm}\.\=\alpha_{\rm mm}^{\rm re})\.\Re(\_H^{\rm loc}_{\rm t}) +
  j(\=I_{\rm t}-\=\beta_{\rm mm}\.\=\alpha_{\rm mm}^{\rm im})\.\Im(\_H^{\rm
    loc}_{\rm t}) \\- \=\beta_{\rm mm}\.\=\alpha_{\rm me}^{\rm
    re}\.\Re(\_E^{\rm loc}_{\rm
    t}) - j\=\beta_{\rm mm}\.\=\alpha_{\rm me}^{\rm im}\.\Im(\_E^{\rm
    loc}_{\rm t}),
\end{multline}
where $\=I_{\rm t}$ is the unit dyadic in the plane of the
array. Respectively,
\begin{multline}
  \l{einc2re} \Re(\_E^{\rm inc}_{\rm t}) = \Re(\=I_{\rm t} - \=\beta_{\rm
    ee}\.\=\alpha_{\rm ee}^{\rm re})\.\Re(\_E^{\rm loc}_{\rm t}) +
  \Im(\=\beta_{\rm ee}\.\=\alpha_{\rm ee}^{\rm im})\.\Im(\_E^{\rm
    loc}_{\rm t}) \\- \Re(\=\beta_{\rm ee}\.\=\alpha_{\rm em}^{\rm
    re})\.\Re(\_H^{\rm loc}_{\rm t}) + \Im(\=\beta_{\rm
    ee}\.\=\alpha_{\rm em}^{\rm im})\.\Im(\_H^{\rm loc}_{\rm t}),
\end{multline}
\begin{multline}
  \l{einc2im} \Im(\_E^{\rm inc}_{\rm t}) = -\Im(\=\beta_{\rm
    ee}\.\=\alpha_{\rm ee}^{\rm re})\.\Re(\_E^{\rm loc}_{\rm t}) +
  \Re(\=I_{\rm t}-\=\beta_{\rm ee}\.\=\alpha_{\rm ee}^{\rm im})\.\Im(\_E^{\rm
    loc}_{\rm t}) \\- \Im(\=\beta_{\rm ee}\.\=\alpha_{\rm em}^{\rm
    re})\.\Re(\_H^{\rm loc}_{\rm
    t}) - \Re(\=\beta_{\rm ee}\.\=\alpha_{\rm em}^{\rm im})\.\Im(\_H^{\rm
    loc}_{\rm t}),
\end{multline}
\begin{multline}
  \l{hinc2re} \Re(\_H^{\rm inc}_{\rm t}) = \Re(\=I_{\rm t} - \=\beta_{\rm
    mm}\.\=\alpha_{\rm mm}^{\rm re})\.\Re(\_H^{\rm loc}_{\rm t}) +
  \Im(\=\beta_{\rm mm}\.\=\alpha_{\rm mm}^{\rm im})\.\Im(\_H^{\rm
    loc}_{\rm t}) \\- \Re(\=\beta_{\rm mm}\.\=\alpha_{\rm me}^{\rm
    re})\.\Re(\_E^{\rm loc}_{\rm t}) + \Im(\=\beta_{\rm
    mm}\.\=\alpha_{\rm me}^{\rm im})\.\Im(\_E^{\rm loc}_{\rm t}),
\end{multline}
\begin{multline}
  \l{hinc2im} \Im(\_H^{\rm inc}_{\rm t}) = -\Im(\=\beta_{\rm
    mm}\.\=\alpha_{\rm mm}^{\rm re})\.\Re(\_H^{\rm loc}_{\rm t}) +
  \Re(\=I_{\rm t}-\=\beta_{\rm mm}\.\=\alpha_{\rm mm}^{\rm im})\.\Im(\_H^{\rm
    loc}_{\rm t}) \\- \Im(\=\beta_{\rm mm}\.\=\alpha_{\rm me}^{\rm
    re})\.\Re(\_E^{\rm loc}_{\rm
    t}) - \Re(\=\beta_{\rm mm}\.\=\alpha_{\rm me}^{\rm im})\.\Im(\_E^{\rm
    loc}_{\rm t}).
\end{multline}
These expressions can be substituted into \r{JmEinc}--\r{JeHinc}
from which one obtains a set of dyadic relations for the
polarizabilities assuming that the four components of the local
fields $\Re(\_E^{\rm loc}_{\rm t}(x,y))$, $\Re(\_H^{\rm loc}_{\rm
t}(x,y))$, $\Im(\_E^{\rm loc}_{\rm t}(x,y))$, and $\Im(\_H^{\rm
loc}_{\rm t}(x,y))$ are independent. Doing so we obtain the
following relations:
\begin{align}
\l{condalpha1}
\_z_0\x\=\alpha_{\rm me}^{\rm re} &= 2j\Im(\=\beta_{\rm
  ee}\.\=\alpha_{\rm ee}^{\rm re})-2\=R\.\Re(\=I_{\rm t}-\=\beta_{\rm
  ee}\.\=\alpha_{\rm ee}^{\rm re}),\\
j\_z_0\x\=\alpha_{\rm me}^{\rm im} &= -2j\Re(\=I_{\rm t}-\=\beta_{\rm
  ee}\.\=\alpha_{\rm ee}^{\rm im})-2\=R\.\Im(\=\beta_{\rm
  ee}\.\=\alpha_{\rm ee}^{\rm im}),\\
\_z_0\x\=\alpha_{\rm mm}^{\rm re} &= 2j\Im(\=\beta_{\rm
  ee}\.\=\alpha_{\rm em}^{\rm re})+2\=R\.\Re(\=\beta_{\rm
  ee}\.\=\alpha_{\rm em}^{\rm re}),\\
j\_z_0\x\=\alpha_{\rm mm}^{\rm im} &= 2j\Re(\=\beta_{\rm
  ee}\.\=\alpha_{\rm em}^{\rm im})-2\=R\.\Im(\=\beta_{\rm
  ee}\.\=\alpha_{\rm em}^{\rm im}),\\
\_z_0\x\=\alpha_{\rm ee}^{\rm re} &= -2j\Im(\=\beta_{\rm
  mm}\.\=\alpha_{\rm me}^{\rm re})+2\=R\.\Re(\=\beta_{\rm
  mm}\.\=\alpha_{\rm me}^{\rm re}),\\
j\_z_0\x\=\alpha_{\rm ee}^{\rm im} &= -2j\Re(\=\beta_{\rm
  mm}\.\=\alpha_{\rm me}^{\rm im})-2\=R\.\Im(\=\beta_{\rm
  mm}\.\=\alpha_{\rm me}^{\rm im}),\\
\_z_0\x\=\alpha_{\rm em}^{\rm re} &= -2j\Im(\=\beta_{\rm
  mm}\.\=\alpha_{\rm mm}^{\rm re})-2\=R\.\Re(\=I_{\rm t}-\=\beta_{\rm
  mm}\.\=\alpha_{\rm mm}^{\rm re}),\\
\l{condalpha2}
j\_z_0\x\=\alpha_{\rm em}^{\rm im} &= 2j\Re(\=I_{\rm t}-\=\beta_{\rm
  mm}\.\=\alpha_{\rm mm}^{\rm im})-2\=R\.\Im(\=\beta_{\rm
  mm}\.\=\alpha_{\rm mm}^{\rm im}),
\end{align}
where $\=R$ is the dyadic reflection coefficient defined in terms of
$R_{\rm TM,TE}$ as \e \=R =
R_{\rm TE}{\_z_0\_z_0\xx\_k_{\rm t}\_k_{\rm t}\over k_{\rm t}^2} +
R_{\rm TM}{\_k_{\rm t}\_k_{\rm t}\over k_{\rm t}^2}  \f
(for the definition of the double cross product and other dyadic algebra rules see, e.g.,
Ref.~\onlinecite{Lindell}). One may
notice that because $\=R$ in the above relations is either zero or
purely imaginary: $(\=R)^*=-\=R$, it follows that $\Re(\=\alpha_{\rm
  ee,mm,em,me}^{\rm re}) = \Im(\=\alpha_{\rm ee,mm,em,me}^{\rm im}) =
0$. Therefore, the equations \r{condalpha1}--\r{condalpha2} can
be also written as
\begin{align}
\l{eqalpha1}
\_z_0\x\=\alpha_{\rm me}^{\rm re} &= 2\left[\Re(\=\beta_{\rm
    ee})+j\=R\.\Im(\=\beta_{\rm ee})\right]\.\=\alpha_{\rm ee}^{\rm
  re}-2\=R,\\
\_z_0\x\=\alpha_{\rm me}^{\rm im} &= 2\left[\Re(\=\beta_{\rm
    ee})+j\=R\.\Im(\=\beta_{\rm ee})\right]\.\=\alpha_{\rm ee}^{\rm
  im}-2\=I_{\rm t},\\
\_z_0\x\=\alpha_{\rm mm}^{\rm re} &= 2\left[\Re(\=\beta_{\rm
    ee})+j\=R\.\Im(\=\beta_{\rm ee})\right]\.\=\alpha_{\rm em}^{\rm
  re},\\
\_z_0\x\=\alpha_{\rm mm}^{\rm im} &= 2\left[\Re(\=\beta_{\rm
    ee})+j\=R\.\Im(\=\beta_{\rm ee})\right]\.\=\alpha_{\rm em}^{\rm
  im},
\end{align}
\begin{align}
\_z_0\x\=\alpha_{\rm ee}^{\rm re} &= -2\left[\Re(\=\beta_{\rm
    mm})-j\=R\.\Im(\=\beta_{\rm mm})\right]\.\=\alpha_{\rm me}^{\rm
  re},\\
\_z_0\x\=\alpha_{\rm ee}^{\rm im} &= -2\left[\Re(\=\beta_{\rm
    mm})-j\=R\.\Im(\=\beta_{\rm mm})\right]\.\=\alpha_{\rm me}^{\rm
  im},\\
\_z_0\x\=\alpha_{\rm em}^{\rm re} &= -2\left[\Re(\=\beta_{\rm
    mm})-j\=R\.\Im(\=\beta_{\rm mm})\right]\.\=\alpha_{\rm mm}^{\rm
  re}-2\=R,\\
\l{eqalpha2}
\_z_0\x\=\alpha_{\rm em}^{\rm im} &= -2\left[\Re(\=\beta_{\rm
    mm})-j\=R\.\Im(\=\beta_{\rm mm})\right]\.\=\alpha_{\rm mm}^{\rm
  im}+2\=I_{\rm t}.
\end{align}

Let us consider first the propagating part of the spectrum. For such
waves, $\=R = 0$, and the above relations simplify. Also, we can
write $\=\beta_{\rm ee} = \eta_0\=\beta$, and $\=\beta_{\rm mm} =
\eta_0^{-1}\=\beta$, where $\=\beta$ is the dimensionless
interaction dyadic which, by duality, is the same for the electric
and magnetic currents as they are due to the electric and magnetic
dipole moments that belong to the same particles in the array.  The
solution of the system of dyadic equations
\r{eqalpha1}--\r{eqalpha2} in the case of $\=R=0$ is
\begin{align}
  &\=\alpha_{\rm ee,mm,em,me}^{\rm re} = 0,\\
\l{sol1}
  &\=\alpha_{\rm me}^{\rm im} = -\=\alpha_{\rm em}^{\rm im}=
  2\left[\=I_{\rm
      t}+4\left(\_z_0\x\Re(\=\beta)\right)^2\right]^{-1}\.(\_z_0\x\=I_{\rm
    t}),\\
\l{sol2}
  &\eta_0\=\alpha_{\rm ee}^{\rm im} = \eta_0^{-1}\=\alpha_{\rm
    mm}^{\rm im} = 4\_z_0\x\Re(\=\beta)\.\left[\=I_{\rm
      t}+4\left(\_z_0\x\Re(\=\beta)\right)^2\right]^{-1}\.(\_z_0\x\=I_{\rm
    t}).
\end{align}
It can be shown~\cite{Maslovski_AEU,Tretyakov_impedanceBC} that the real part of the interaction
dyadic $\=\beta$ for a planar array verifies \e
\l{k3na6pi}\Re(\=\beta) =-{1\over
  2\cos\theta}{\_z_0\_z_0\xx\_k_{\rm t}\_k_{\rm t}\over k_{\rm t}^2}
-{\cos\theta\over 2}{\_k_{\rm t}\_k_{\rm t}\over k_{\rm t}^2} +
{k_0^2A_0\over 6\pi}\=I_{\rm t}, \f where $\theta$ is the angle of
incidence: $\cos\theta = \sqrt{1-k^2_{\rm t}/k_0^2}$, and $A_0$ is the
unit cell area. This result holds for arrays with arbitrary unit cell
geometries, provided that the arrays do not produce higher-order
diffraction lobes.

It is quite interesting that the imaginary part of the interaction
constant that contains the information about the microstructure of
the array has completely disappeared from the above solution. The
imaginary part of the interaction constant does not contribute in
this case because the induced currents $\_J_{\rm e}$ and $\_J_{\rm
m}$ are always imaginary, and the respective additions to the
interaction field $j\Im(\=\beta_{\rm ee})\.\_J_{\rm e}$ and
$j\Im(\=\beta_{\rm mm})\.\_J_{\rm
  m}$ are real-valued, to which the particles do not react. Thus, the
interaction of the particles in the array is irrelevant in the
considered case, and each particle radiates effectively as in free
space.

We may substitute \r{k3na6pi} into \r{sol1}--\r{sol2}, taking into
account that \e \left(\_z_0\x\Re(\=\beta)\right)^2 = \left[-{1\over
4} -
  \left({k_0^2A_0\over 6\pi}\right)^2 + {k_0^2A_0\over
    12\pi}\left(\cos\theta+{1\over \cos\theta}\right)\right]\=I_{\rm t},
\f
and obtain
\begin{align}
\l{finsol1}
  &\=\alpha_{\rm me}^{\rm im} = -\=\alpha_{\rm em}^{\rm im} =
  {6\pi\over k_0^2 A_0}\left(\cos\theta+{1\over \cos\theta}-{k_0^2A_0\over 3\pi}\right)^{-1}(\_z_0\x\=I_{\rm t}),\\
\l{finsol2}
  &\eta_0\=\alpha_{\rm ee}^{\rm im} = \eta_0^{-1}\=\alpha_{\rm
    mm}^{\rm im} =
  -{12\pi\over k_0^2 A_0}\left(\cos\theta+{1\over \cos\theta}-{k_0^2A_0\over 3\pi}\right)^{-1}(\_z_0\_z_0\xx\Re(\=\beta)).
\end{align}
For practical purposes, considering the phase conjugation of paraxial beams in
dense arrays, we may approximate the above relations as
\begin{align}
\l{appsol1}
  &\=\alpha_{\rm me}^{\rm im} = -\=\alpha_{\rm em}^{\rm im} \approx
  {3\pi\over k_0^2 A_0}(\_z_0\x\=I_{\rm t}),\\
\l{appsol2}
  &\eta_0\=\alpha_{\rm ee}^{\rm im} = \eta_0^{-1}\=\alpha_{\rm
    mm}^{\rm im} \approx {3\pi\over k_0^2 A_0}\=I_{\rm t}.
\end{align}
Because of the form of the relations \r{Einc1}--\r{Hinc2}, the
obtained exact solutions \r{sol1}--\r{sol2} and their approximations
\r{appsol1}--\r{appsol2} are valid for arbitrary plane waves
incident from the half space $z < 0$. The solution for the case of
incidence from the half space $z
> 0$ is obtained by replacing $\_z_0$ with $-\_z_0$ in
\r{sol1}--\r{sol2} and \r{appsol1}--\r{appsol2}, which changes signs
of $\=\alpha_{\rm em,me}^{\rm im}$.

From the above results we see that the particle must be
``invisible'' for the real-valued electric and magnetic fields,
while the polarizabilities of the particle to the imaginary-valued
fields must be such that the electric and magnetic currents form
Huygens pairs that absorb the incident wave and produce the
phase conjugated wave.

For the evanescent waves $|R_{\rm TM,TE}|\rightarrow\infty$, therefore,
it is convenient to multiply the equations \r{eqalpha1}--\r{eqalpha2}
by $\=R^{-1}$ from the left. Then, in the limit $\=R^{-1}\rightarrow
0$ the following solution of the system \r{eqalpha1}--\r{eqalpha2} can
be immediately found:
\begin{align}
\l{evan1}
&\eta_0\=\alpha_{\rm ee}^{\rm re}=\eta_0^{-1}\=\alpha_{\rm mm}^{\rm
  re}=-j\left[\Im(\=\beta)\right]^{-1},\\
\l{evan2}
&\=\alpha_{\rm ee,mm}^{\rm im} = \=\alpha_{\rm em,me}^{\rm re,im} = 0.
\end{align}
The same solution can be also obtained with a more accurate
treatment. Let us introduce the notations $\=C_{\rm e} =
\Re(\=\beta_{\rm ee})+j\=R\.\Im(\=\beta_{\rm ee})$ and $\=C_{\rm m} =
\Re(\=\beta_{\rm mm})-j\=R\.\Im(\=\beta_{\rm mm})$. Then, in these
notations, we may, for example, write the solution for $\=\alpha_{\rm
  ee}^{\rm re}$ as
\begin{multline}
  \=\alpha_{\rm ee}^{\rm re} = 4\left[\=I_{\rm t}+4(\_z_0\x\=C_{\rm
    m})\.(\_z_0\x\=C_{\rm e})\right]^{-1}\.(\_z_0\x\=C_{\rm
    m})\.(\_z_0\x\=R)\\
= \left[\=I_{\rm t}+{1\over 4}(\_z_0\x\=C_{\rm
    e})^{-1}\.(\_z_0\x\=C_{\rm m})^{-1}\right]^{-1}\.(\_z_0\x\=C_{\rm
  e})^{-1}\.(\_z_0\x\=R)\\
 = \=C_{\rm e}^{-1}\.\=R + {\cal O}\big(\=R^{-2}\big).
\end{multline}
Next,
\begin{multline}
\=C_{\rm e}^{-1}\.\=R = -j\left[\Im(\=\beta_{\rm ee})\right]^{-1}\.\left[\=I_{\rm
  t}-j\=R^{-1}\.\Re(\=\beta_{\rm ee})\.\left(\Im(\=\beta_{\rm
  ee})\right)^{-1}\right]^{-1} \\
= -j\left[\Im(\=\beta_{\rm ee})\right]^{-1} + {\cal O}\big(\=R^{-1}\big),
\end{multline}
which leads to~\r{evan1}.

From~\r{evan1}--\r{evan2} we conclude that to conjugate the evanescent
part of the spectrum the inclusions must be ``invisible'' to the
imaginary part of the electric and magnetic fields. The inclusions do
not have to be bi-anisotropic in this case. The particles are purely reactive and
their reactance should compensate the reactance due to particle interactions, creating a resonant structure.

From a physical point of view, condition \r{evan1} can be understood
as a condition for a surface polariton resonance at the array
surface. Indeed, for particles with the
polarizabilities~\r{evan1}--\r{evan2} we may write $\_J_{\rm e} =
\=\alpha_{\rm ee}^{\rm re}\.\Re(\_E^{\rm loc})$. From the other
hand, $\_E^{\rm loc} = \_E^{\rm inc} + \=\beta_{\rm ee}\.\_J_{\rm
  e}$. Therefore,
\begin{equation}
\_J_{\rm e} =
\=\alpha_{\rm ee}^{\rm re}\.\Re(\_E^{\rm inc} + \=\beta_{\rm ee}
\.\_J_{\rm e}) =
\=\alpha_{\rm ee}^{\rm re}\.\Re(\_E^{\rm inc}) + j\=\alpha_{\rm
  ee}^{\rm re}\.\Im(\=\beta_{\rm ee})\.\_J_{\rm e},
\end{equation}
because both $\=\alpha_{\rm ee}^{\rm re}$ and $\_J_{\rm e}$ are purely
imaginary. From here \e \_J_{\rm e} = \left[\=I_{\rm t} -
  j\=\alpha_{\rm ee}^{\rm re}\.\Im(\=\beta_{\rm
    ee})\right]^{-1}\.\=\alpha_{\rm ee}^{\rm re}\.\Re(\_E^{\rm inc}),
\f and we see that $\_J_{\rm e}\rightarrow\infty$ when condition
\r{evan1} is fulfilled. A similar resonance is responsible for the
enhancement of the evanescent waves in a pair of linear
plasmon-polariton resonant grids studied in previous works.\cite{Maslovski_grid,Marques_SRR,Alitalo_cylinder,Alitalo_nanospheres}

\subsection{Electromagnetic properties of the particles forming phase
  conjugating sheets}

Although the principle of operation of the field-conjugating perfect
lens is the nonlinear operation of complex conjugation of
electromagnetic fields, it is interesting to observe that the
particles which perform this operation are characterized by linear
polarizabilities with respect to the {\itshape real or imaginary
  parts} of the fields. The nonlinear nature of the particles is thus
only in their selective sensitivity to either real or imaginary parts
of the complex amplitude of the local fields.

Considering the particle response to the real or imaginary field
components separately, we may apply the theory of usual linear
bi-anisotropic particles. For the particles which react to the
imaginary parts of the field (the particles excited by the propagating
part of the spatial spectrum), we rewrite relations \r{elepolar},
\r{magpolar}, and \r{appsol1} in terms of the induced electric and
magnetic dipole moments of individual particles $\_p_{\rm e,m}$ and
the local fields:
$$ \_p_{\rm e}=\left(-j{3\pi\epsilon_0\over k_0^3}\right)j\Im(\_E^{\rm
    loc}_{\rm t})+\eta_0\left(j{3\pi\epsilon_0\over
      k_0^3}\right)\_z_0\x j\Im(\_H^{\rm loc}_{\rm
    t})$$ \e =\=a_{\rm ee}\.j\Im(\_E^{\rm
    loc}_{\rm t})+\eta_0\=a_{\rm em}\.j\Im(\_H^{\rm loc}_{\rm
    t}), \l{pe}\f
$$ \_p_{\rm m}=\left(-j{3\pi\mu_0\over k_0^3}\right)j\Im(\_H^{\rm
    loc}_{\rm t})-{1\over \eta_0}\left(j{3\pi\mu_0\over
      k_0^3}\right)\_z_0\x j\Im(\_E^{\rm loc}_{\rm
    t})$$ \e =\=a_{\rm mm}\.j\Im(\_H^{\rm
    loc}_{\rm t})+{1\over \eta_0}\=a_{\rm me}\.j\Im(\_E^{\rm loc}_{\rm
    t}) \l{pm}\f
(the surface current densities are related to the dipole moments of
individual particles as $\_J_{\rm e,m}=j\omega \_p_{\rm e,m}/A_0$).

These relations show that the particles reacting to the propagating
part of the spectrum are bi-anisotropic and nonreciprocal. The
magnetoelectric coupling is due to nonreciprocity only (no
magnetoelectric coupling due to reciprocal spatial dispersion
effects), because the coupling dyadics satisfy \e \=a_{\rm
  em}=\=a_{\rm me}^T. \f Furthermore, because these dyadics are
antisymmetric ($\=a_{\rm em}=-\=a_{\rm em}^T$, $\=a_{\rm me}=-\=a_{\rm
  me}^T$), materials formed by particles of this type belong to the
class of moving media.\cite{interact,biama}

The polarizabilities of lossless bi-anisotropic particles satisfy the
following conditions (e.g., Ref.~\onlinecite{biama}): \e \=a_{\rm ee}=\=a_{\rm
  ee}^\dagger, \quad \=a_{\rm mm}=\=a_{\rm mm}^\dagger, \quad \=a_{\rm
  em}=\=a_{\rm me}^\dagger,\f where $\dagger $ denotes the Hermitian
conjugation operation. Obviously, the inclusions with the
polarizabilities \r{pe} and \r{pm} have the opposite property of being
purely passive or active (there is no stored electromagnetic energy in
their near fields), because they satisfy the opposite conditions: \e
\=a_{\rm ee}=-\=a_{\rm ee}^\dagger, \quad \=a_{\rm mm}=-\=a_{\rm
  mm}^\dagger, \quad \=a_{\rm em}=-\=a_{\rm me}^\dagger.\f The power
extracted from the local fields by one pair of the particles reads \e
P={1\over 2}{\rm Re}\{\_J_{\rm e}^*\cdot \_E^{\rm loc}+\_J_{\rm
  m}^*\cdot \_H^{\rm loc}\}A_0.\f Assuming paraxial wave propagation,
we can use the polarizability expressions \r{appsol1} and \r{appsol2}
and assume that the electric and magnetic local fields are related by
the free-space wave impedance $\eta_0$. In this approximation we find,
for the case of plane wave incidence from the half space $z < 0$, \e
P= {1\over 2}{3\pi \over k_0^2 } {4\over \eta_0}[{\rm Im}(E^{\rm
  loc})]^2={6\pi \over k_0^2 \eta_0}[{\rm Im}(E^{\rm loc})]^2 ={6\pi
  \eta_0 \over k_0^2 }[{\rm Im}(H^{\rm loc})]^2.\f As is clear from
this result, each of the polarizability components in \r{appsol1} and
\r{appsol2} brings equal contributions to the extracted power. Noting
that the induced dipole moments of ideal absorption-free dipole
scatters read (at the resonance) \e \_p_{\rm e0}=\left(-j{6\pi\epsilon_0\over
    k_0^3}\right)\_E^{\rm loc}, \qquad \_p_{\rm
  m0}=\left(-j{6\pi\mu_0\over k_0^3}\right)\_H^{\rm loc},\f we see
that a pair of such ideal dipoles would extract from the fields
exactly the same amount of power as our phase conjugating particles
(when the complex amplitude of the local field is purely imaginary at
the point of the particle). Thus, we can conclude that the particles
described by \r{appsol1} and \r{appsol2} actually do not absorb
power. They act as ideal absorption-free particles, which receive
power from the incident field and re-radiate the same amount of power,
creating phase conjugated waves of the same intensity as the incident
propagating waves.

It is easy to check that the same particles with the
polarizabilities \r{pe}--\r{pm} do not react to the plane waves
incident from the half space $z > 0$. Indeed, the sign of the
magnetic field in these waves is opposite, therefore, the
contributions due to $\_E^{\rm loc}$ and $\_H^{\rm loc}$ to the
electric dipole moment $\_p_{\rm e}$ and the magnetic dipole moment
$\_p_{\rm m}$ compensate each other in Eqs.~\r{pe}--\r{pm}, so that
$\_p_{\rm e} = \_p_{\rm m} = 0$ under such excitation. As was
mentioned in Section~\ref{genreq}, to conjugate the waves incident
from the half space $z > 0$, one must change the signs of the
magnetoelectric coupling terms in \r{pe}--\r{pm}. Physically, this
requires another array of inclusions with slightly different
topology (more details in the next section). Fortunately, the
particles in the two arrays do not interact, so that in practice it
is possible to combine the two types of particles in a single plane,
for example, in a chess board-like structure.

\section{\label{secdesign}Design of the phase conjugating
  bi-anisotropic inclusions at microwaves}

\subsection{\label{desprop}Case of propagating waves}

To approach the design of nonlinear phase conjugating inclusions at
microwaves one may start from the ideas behind the well-known omega
particle.\cite{saa,1993} An omega particle is a combination of a
short dipole antenna and a small loop antenna. The particle is
planar, so that both the dipole and the loop may be printed on a
single side of a printed-circuit board. In the most common
variant of this linear and reciprocal inclusion the dipole is
directly connected to the loop. The nonlinearity and nonreciprocity,
thus, can be achieved if one inserts a nonlinear and nonreciprocal
four-pole network between the two antennas.

To identify what kind of network one may need, let us first briefly
analyze how the linear omega particle reacts to the local electric and
magnetic fields. We consider the case when the particle lies in the
$xz$-plane with the dipole antenna oriented along the $x$-axis. In
this geometry, the dipole reacts to the $x$-component of the electric
field, $E_x^{\rm loc}$, and the loop reacts to the $y$-component of the
magnetic field, $H_y^{\rm loc}$.

The electromotive force (EMF) induced by the local field in the
dipole can be written as \e {\cal E}_{\rm dip} = h_{\rm dip}E_x^{\rm
loc}, \l{EMFdip}\f where $h_{\rm dip}$ is the effective height of
the dipole antenna. For a short dipole of the total length $2l$, the
effective height is $h_{\rm
  dip} = l$. Respectively, the EMF induced in the loop by the magnetic
field reads \e {\cal E}_{\rm loop} = -j\o \M_0 S H_y^{\rm loc},
\l{EMFloop}\f where $S=\pi r^2$ is the area of the loop. Under a
normal plane wave incidence, $H_y^{\rm loc} = \pm E_x^{\rm
loc}/\eta_0$ (the two signs are for the two possible directions of
incidence), therefore, \e {\cal E}_{\rm
  loop} = \mp j (k_0 S) E_x^{\rm loc} = \mp j h_{\rm loop} E_x^{\rm
  loc}, \f where we have introduced the effective height of a small
loop antenna $h_{\rm loop} = k_0 S$. One may notice that the EMF
induced in the loop is in quadrature with respect to the EMF in the
dipole. Therefore, when the two antennas are directly connected,
these EMFs add up, but never fully compensate (or fully complement)
each other. In fact, in the most interesting case when $h_{\rm loop}
= h_{\rm dip} = l$, the total EMF induced in the antennas is \e
{\cal E}_{\rm tot} = (1\mp j)lE_x^{\rm loc}.\f Respectively, the
induced current at the point where the dipole connects to the loop
is \e I_{\rm dip} = {(1\mp j)l E_x^{\rm loc}\over Z_{\rm dip} +
Z_{\rm
    loop}}, \f where $Z_{\rm dip}$ is the input impedance of the
dipole and $Z_{\rm loop}$ is the input impedance of the loop. The
induced electric and magnetic dipole moments are proportional to this
current:
\begin{align}
\l{pelin}
  p_{{\rm e},x} &= {I_{\rm dip}l\over j\o} = -j{(1\mp j)l^2
    E_x^{\rm loc}\over \o(Z_{\rm dip} + Z_{\rm loop})},\\
\l{pmlin}
  p_{{\rm m},y} &= \M_0I_{\rm dip}S = \eta_0{(1\mp j)l^2 E_x^{\rm
      loc}\over \o(Z_{\rm dip} + Z_{\rm loop})},
\end{align}
where we use the fact that $S = l/k_0$ if $h_{\rm loop}
= h_{\rm dip} = l$. When the particle is at
resonance, $Z_{\rm dip} + Z_{\rm loop} = 2R_{\rm rad} + R_{\rm loss}$,
where $R_{\rm rad}={\ds\eta_0k_0^2l^2\over \ds6\pi}$ is the radiation
resistance of a short dipole antenna (when $h_{\rm loop} = h_{\rm
  dip}$ both antennas have the same radiation resistance) and
$R_{\rm loss}$ corresponds to the ohmic loss in metal, which we may
neglect. From these relations we see that the induced dipole moments in the
linear omega particle are in quadrature.

However, from \r{pe}--\r{pm} it follows, first, that in the nonlinear
particle which we want to design the contributions due to $\_E^{\rm loc}$
and $\_H^{\rm loc}$ in the expressions for both dipole moments must be
in phase (for the wave incident from $z < 0$), and, second, the
induced electric and magnetic moments themselves must be also in
phase. In a symbolic language, we may say that the four-pole
network that we insert between the two antennas must act in such a way
that $\mp j$ in \r{pelin}--\r{pmlin} is replaced with $\pm 1$, and
$\eta_0$ in~\r{pmlin} with $-j\eta_0$. As will be seen in a few
moments, this can be achieved with microwave nonlinear circuits
known as {\em balanced modulators} (BM).

\begin{figure}
\centering
\epsfig{file=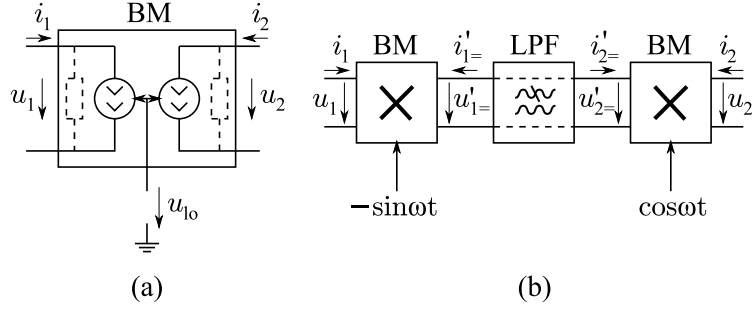,width=0.55\textwidth}
\caption{\label{bmcirc} (a) Equivalent circuit of an idealized
  balanced modulator (BM). (b) A network composed of two balanced
  modulators connected through a low-pass filter (LPF).}
\end{figure}

A balanced modulator is a three-port device such that there are two
ports that may both serve as input and output (the ports are
exchangeable due to the symmetry of the circuit; that is one of the
reasons why the circuit is called {\it balanced} modulator), and the
third port to which a voltage from a local oscillator is applied. The
function that the BM performs is a multiplication of the signal applied to
one of its input ports and the signal of the local oscillator.
One may represent an idealized BM with an equivalent circuit shown in
Fig.~\ref{bmcirc}(a). The controllable current sources in the circuit
depend on the instantaneous voltages at the ports as follows:
\begin{align}
i_1(t) = K_{12} u_2(t) u_{\rm lo}(t),\\
i_2(t) = K_{21} u_1(t) u_{\rm lo}(t),
\end{align}
where $u_{\rm lo}(t)$ is the voltage at the local oscillator port and
$u_{1,2}(t)$ and $i_{1,2}(t)$ are the voltages at the other two
ports. We need power conserving BMs which do not absorb or store power
that is delivered to ports 1 and 2, hence, $u_1(t) i_1(t) + u_2(t)
i_2(t) = 0$. From here, $K_{12}= -K_{21}$.  The input and output
resistances of the BM shown in the figure with dashed lines are
assumed to be very large and are not taken into account.

Consider now the network depicted in Fig.~\ref{bmcirc}(b). In this
network we have connected two BMs through a low-pass filter (LPF). One
may imagine this LPF as a $\Pi$-type $CLC$-filter with an inductor in
a series branch and two capacitors in the parallel branches. For us,
however, the only thing that is important here is that this LPF freely
passes through the direct-current (DC) component and blocks all
high-frequency components (the DC path through the filter is shown
with dashed lines).

The whole network is designed to operate with
signals at the frequency $\o = 2\pi f_0 = 2\pi/T_0$, therefore, we may
represent the voltages $u_{1,2}(t)$ as \e u_{1,2}(t) = U_{1,2}^{\rm
  re}\cos\o t - U_{1,2}^{\rm im}\sin\o t.  \f We apply the local
oscillator signal at the frequency $\o$ and the phase $\varphi=\pi/2$
to the first BM: $u_{\rm lo,1} = -\sin\o t$, and another signal at the
same frequency and $\varphi = 0$ to the second BM: $u_{\rm lo,2} =
\cos\o t$. Therefore, we may write for the DC currents $i'_{1,2=}$
(here $\langle\ldots\rangle_{T_0}$ denotes averaging over a period):
\begin{align}
i'_{1=} &= K_{21}\left\langle(U_{1}^{\rm   re}\cos\o t - U_{1}^{\rm
  im}\sin\o t)(-\sin\o t)\right\rangle_{T_0} = {1\over 2}K_{21}U_{1}^{\rm
im},\\
i'_{2=} &= K_{12}\left\langle(U_{2}^{\rm   re}\cos\o t - U_{2}^{\rm
  im}\sin\o t)\cos\o t\right\rangle_{T_0} = {1\over 2}K_{12}U_{2}^{\rm
re}.
\end{align}
But as is dictated by the topology of the network, $i'_{1=} =
-i'_{2=}$, therefore, \e U_{1}^{\rm im} = U_{2}^{\rm re}. \l{volt}\f
Next, we express the high-frequency currents at the ports 1 and 2:
\begin{align}
i_1(t) &= (K_{12}u'_{1=})(-\sin\o t) = -I_1^{\rm im}\sin\o t,\\
i_2(t) &= (K_{21}u'_{2=})\cos\o t = I_2^{\rm re}\cos\o t,
\end{align}
where $I_1^{\rm im}=K_{12}u'_{1=}$ and $I_2^{\rm re}=K_{21}u'_{1=}$.
But again, from the topology of the network, the DC voltages satisfy
$u'_{1=} = u'_{2=}$, therefore,
\e
I_1^{\rm im} = -I_2^{\rm re}.
\l{curr}
\f
Notice also that always $I_1^{\rm re} = I_2^{\rm im} = 0$.

\begin{figure}
\centering
\epsfig{file=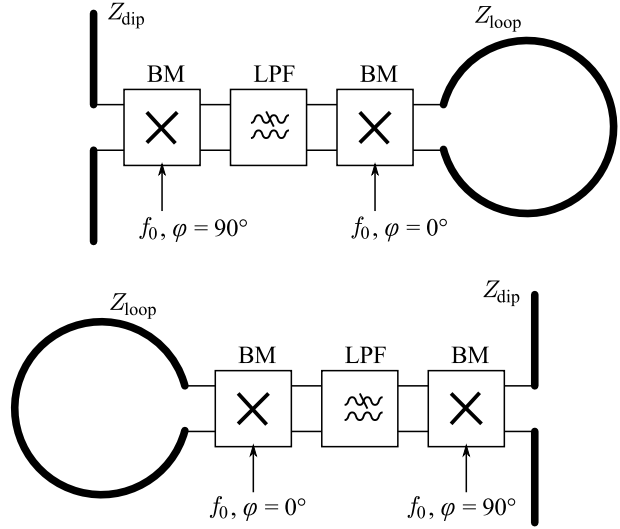,width=0.45\textwidth}
\caption{\label{design} The topology of the two types of phase
  conjugating bi-anisotropic inclusions for the phase
  conjugating surface operating at the frequency $f_0$. Each particle
  consists of a short dipole antenna with the impedance $Z_{\rm dip}$
  and a small loop antenna with the impedance $Z_{\rm loop}$. The two
  antennas are interconnected through a nonlinear and nonreciprocal
  network composed of two balanced modulators (BM) and a low-pass
  filter (LPF). The mixers are fed by a local oscillator with two
  signal outputs in quadrature $\varphi = 0^\circ$ and $\varphi =
  90^\circ$. See the main text for further explanations.}
\end{figure}

Thus, from \r{volt} and \r{curr} it is evident that the considered
network operates, essentially, as a ``connector'' between the
imaginary current and voltage at the first port, and the real current
and voltage at the second port. The network also enforces zero real
current in the first port and zero imaginary current in the second
port. This is exactly what we need in the design of the phase
conjugating particles, and the corresponding topologies including the
antennas are shown in Fig.~\ref{design}.

In these designs we connect the dipole and loop antennas
to the BM-based network discussed above. The electrical size of the circuit is negligible,
and both antennas are excited by the same local field. Let us analyze the operation
of the variant shown at the top of Fig.~\ref{design}.  First of all, it
is evident that when placed in an arbitrary local field, the EMFs in
both antennas are still given by \r{EMFdip} and \r{EMFloop}. However,
contrary to what happens in a linear omega particle, the real part of
${\cal E}_{\rm dip}$ and the imaginary part of ${\cal E}_{\rm loop}$
will not be able to excite any current in the antennas, because these
components are blocked by the BMs. Therefore, the relevant parts of
the EMFs in the two antennas are (as above, we let $h_{\rm dip} = h_{\rm
  loop} = l$)
\begin{align}
\l{Edip}
{\cal E}_{\rm dip}^{\rm im} &= l\Im(E_x^{\rm loc}),\\
\l{Eloop}
{\cal E}_{\rm loop}^{\rm re} &= \eta_0l\Im(H_y^{\rm loc}).
\end{align}
Analogously, the voltage drops on the reactive parts $X_{\rm dip} =
\Im(Z_{\rm dip})$ and $X_{\rm loop} = \Im(Z_{\rm loop})$ of the input
impedances of the dipole and the loop have no effect as well, as they
are in quadrature with respect to the current flowing through
them. Thus, the only relevant part of the input impedance is the
radiation resistance.

From the topology of the network, $I_{\rm dip}^{\rm im} = -I_1^{\rm im}$ and
$I_{\rm loop}^{\rm re} = I_2^{\rm re}$, therefore, $I_{\rm dip}^{\rm im} = I_{\rm
  loop}^{\rm re}$. Next, because of \r{volt}, the EMFs \r{Edip} and \r{Eloop}
and the rest of the equivalent circuits of the two antennas (only two
$R_{\rm rad}$ remain) appear to be essentially connected in series,
hence,
\e
I_{\rm dip}^{\rm im} = I_{\rm loop}^{\rm re} = {l[\Im(E_x^{\rm loc}) +
  \eta_0\Im(H_y^{\rm loc})]\over 2R_{\rm rad}}.
\f
The complex amplitudes of the electric and magnetic dipole moments,
therefore, read
\begin{align}
p_{{\rm e},x} &= {l(jI_{\rm dip}^{\rm im})\over j\o} = {l^2\over 2\o R_{\rm rad}}[\Im(E_x^{\rm loc}) +
  \eta_0\Im(H_y^{\rm loc})],\\
p_{{\rm m},y} &= \M_0 I_{\rm loop}^{\rm re} S = \eta_0{l^2\over 2\o R_{\rm rad}}[\Im(E_x^{\rm loc}) +
  \eta_0\Im(H_y^{\rm loc})],
\end{align}
which is a particular case of \r{pe}--\r{pm} for the considered
polarization.

As can be readily verified, the second design variant shown at the
bottom of Fig.~\ref{design} has the opposite signs of the
magnetoelectric interaction terms, and, thus, must be used to
conjugate the plane waves incident from the half space $z > 0$.

\subsection{Case of evanescent waves}

To phase-conjugate the evanescent waves, the particles must react to
the real parts of electric and magnetic fields, as follows
from~\r{evan1}--\r{evan2}. In this case the inclusions are simple
electric and magnetic dipoles without magneto-electric
interaction. Therefore, as the base for our design we may choose the
loaded dipole and loop antennas. In what follows, we consider in
detail the case of a linear electric dipole oriented along the
$x$-axis (the magnetic dipole along the $y$-axis may be considered in
a dual manner).

When a short loaded dipole is placed in an electric field, an EMF is
induced in the dipole, with the value given
by~\r{EMFdip}. Respectively, the current induced in the dipole is \e
I_{\rm dip} = {l E_x^{\rm loc}\over Z_{\rm dip} + Z_{\rm load}}, \f
where $Z_{\rm load}$ is the impedance of a bulk load connected to the
dipole. The induced electric dipole moment of the loaded dipole reads
\e p_{{\rm e},x} = {I_{\rm dip}l\over j\o} = -j{l^2 E_x^{\rm loc}\over
  \o(Z_{\rm dip} + Z_{\rm load})}.  \f It is easy to verify that if we
chose the load so that \e Z_{\rm load} = -Z_{\rm dip} + {j\eta_0
  l^2\over A_0}\Im(\beta), \l{loaddip}\f then the condition for the surface
electric current polarizability~\r{evan1} in an array of such
particles will be satisfied, with an exception that the linear
particles will react to both real and imaginary parts of the electric
field.

\begin{figure}
\centering
\epsfig{file=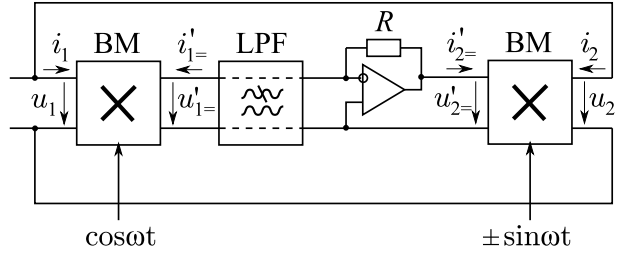,width=0.45\textwidth}
\caption{\label{evannet}A nonlinear single-port network used as a load
  for a short electric dipole. The network is composed of the same
  elements as in Fig.~\ref{bmcirc}, with an additional operational
  amplifier working in the current-to-voltage conversion mode.}
\end{figure}

With the use of BMs we may get rid of the reaction to the imaginary
part of the field and also find a simple way to realize the necessary
loading~\r{loaddip}. Consider the network shown in Fig.~\ref{evannet}.
This network is similar to the ones we considered in
Sec.~\ref{desprop}. It is composed of a pair of BMs and a LPF, with an
additional element in the middle that is an operational amplifier
working in the current-to-voltage conversion mode. We pump the first
BM at the frequency $f_0$ with the phase $\varphi = 0$, and the second
one with the phase $\varphi = \pm\pi/2$ (it will be seen soon why we
consider two signs of the phase). The output of the network is
directly connected to its input, so that the circuit is essentially a
single port device that may be used as an active load. To analyze the
operation of the circuit we first notice that because the voltage at
the input of the operational amplifier is negligibly small,
$u'_{1=}=0$, and, therefore, $i_1 = 0$. Next, assuming that the
voltage at the input of the circuit is $u_1 = U_1^{\rm re}\cos\o t -
U_1^{\rm im}\sin\o t$, $\o = 2\pi f_0 = 2\pi/T_0$, we obtain \e
i'_{1=} = K_{21}\left\langle(U_1^{\rm re}\cos\o t - U_1^{\rm im}\sin\o
  t)\cos\o t\right\rangle_{T_0} = {1\over 2}K_{21}U_1^{\rm re}.\f
Hence, the DC voltage at the output of the operational amplifier is \e
u'_{2=} = i'_{1=}R = {1\over 2}K_{21} R U_1^{\rm re}, \f where $R$ is
the resistor in the feedback loop of the operational
amplifier. Therefore, \e i_2(t) = \mp K_{21}u'_{2=}\sin\o t =
\mp{1\over 2}K_{21}^2 R U_1^{\rm re}\sin\o t = -I_2^{\rm im}\sin\o t,
\f where $I_2^{\rm im} = \pm K_{21}^2R U_1^{\rm re}/2$. This current
is the input current of the whole network, because the input current
of the first BM equals zero. Thus, we have designed a circuit in which
a real input voltage induces an imaginary current, i.e., the circuit
behaves almost as a usual reactance (the plus sign in the expression
for the current corresponds to the capacitance and the minus sign ---
to the inductance), with the difference that the loading circuit is
not sensitive at all to the imaginary input voltage. This is exactly
what we need to realize the nonlinear particles reacting only to the
real part of the electric field. Indeed, to realize the required
loading one has to choose the parameters of the circuit so that \e \pm
{1\over 2}K_{21}^2R = -X_{\rm dip} + {\eta_0 l^2\over A_0}\Im(\beta).
\f It is interesting to note that there is no need to compensate the
real part of the dipole impedance $Z_{\rm dip}$ (the radiation
resistance), because the circuit reacts only to the real part of the
input voltage, and the additional voltage drop on the radiation
resistance $U_R=jI_2^{\rm im}R_{\rm rad}$ is purely imaginary.

\begin{figure}
\centering
\epsfig{file=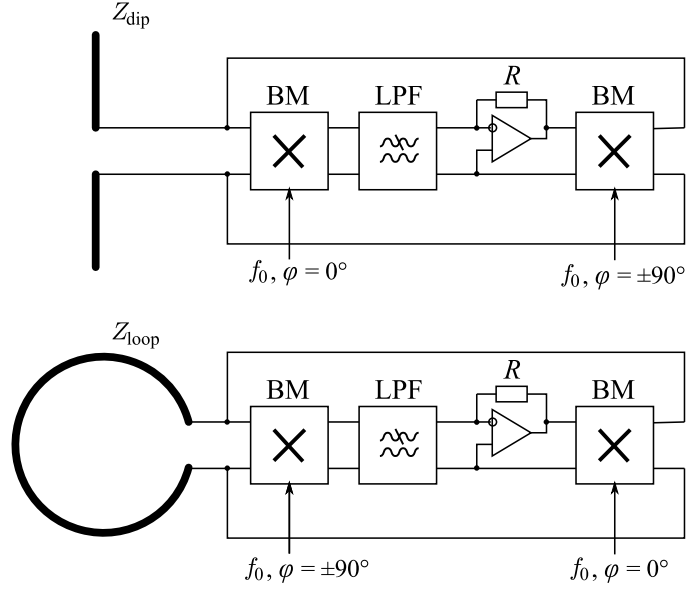,width=0.5\textwidth}
\caption{\label{evantop} The topology of the two particles loaded with
nonlinear circuits designed to operate with the evanescent part of
the spatial spectrum. Top: a short electric dipole loaded with a nonlinear
load. Bottom: a small magnetic loop loaded with a nonlinear load.}
\end{figure}

An example topology of an electric dipole particle with the nonlinear
active load is shown in Fig.~\ref{evantop}, top, and the same for the
magnetic dipole particle in Fig.~\ref{evantop}, bottom (notice the
difference in phases of the local oscillator signals in both
schematics). We would like to stress that the operational amplifier
seen in this schematic works with a DC signal. Thus, its role is {\it
  not} to amplify the evanescent fields of an incoming wave, but just
to provide the necessary {\it reactance} of the loading circuit in
order to tune the structure to a resonance. In turn, the evanescent
modes in this structure are enhanced because of this resonance.

\section{Conclusions}

In this paper, the concept of perfect lensing with a pair of phase
conjugating surfaces introduced by us earlier,\cite{JAP} i.e., a
possibility to achieve optical resolution well below the wavelength
limit without using double-negative materials, has been further
developed. Working as a planar lens, a pair of phase conjugating
sheets is able to focus propagating modes of a source due to the
negative refraction at the interfaces and, in the same time, enhance
the evanescent modes due to surface plasmon-polariton resonances,
i.e., it provides sub-wavelength resolution imaging
indistinguishable from the perfect lens proposed by Pendry, while
not suffering from its known drawbacks.

We have investigated in detail the physics of operation of nonlinear
sheets with the boundary conditions of the form $ {\_E_{\rm
    t}}(\o)|_1=\_E_{\rm t}(\o)^*|_2$, $\_H_{\rm t}(\o)|_1=\_H_{\rm
  t}(\o)^*|_2 $ and have demonstrated that they are, in principle,
physically realizable with devices imposing the necessary relations
between the fields and the equivalent electric and magnetic surface
currents at the phase conjugating boundary.  Namely, we have shown
that the mentioned surface currents must form Huygens sources that
radiate towards a given side of the boundary, negating the fields
incident from the other side and creating complex-conjugated fields in
the corresponding half space.

As a possible realization of such sheets, we have proposed and
considered in this paper arrays of nonlinear and nonreciprocal
bi-anisotropic inclusions reacting differently to the propagating and
evanescent plane waves. At microwaves, the considered design makes use
of balanced modulators (a type of mixers) to provide for the
required nonlinearity and nonreciprocity of the circuit. At optics, a
design utilizing similar principles may become feasible in future as
the field of optical nanocircuits develops further.

As a final note we would like to mention that in such arrays the
enhancement of the evanescent waves is due to a high-quality surface
mode resonance, as in the grids of passive resonant inclusions
considered in Ref.~\onlinecite{Maslovski_grid}. This is in contrast to
Ref.~\onlinecite{Pendry} where the phase conjugating surface must
itself parametrically amplify the fields which requires an
unphysically high conversion efficiency.

\end{document}